\documentclass[12pt]{article}
\usepackage{amssymb}
\usepackage{amsmath}
\usepackage{cite}
\usepackage{axodraw}
\usepackage{hyperref}
\input xy
\xyoption{all}
\usepackage{rotfloat}   

\newcommand{\Z}{\mathbb{Z}}

\def\sqr#1#2{{\vcenter{\vbox{\hrule height.#2pt
            \hbox{\vrule width.#2pt height#1pt \kern#1pt
                  \vrule width.#2pt}\hrule height.#2pt}}}}

\def\square
 {\mathop{\mathchoice{\sqr{12}{15}}{\sqr{9}{12}}
{\sqr{6.3}{9}}{\sqr{4.5}{9}}}}

\def\sqra#1#2#3{{\vcenter{\vbox{\hrule height.#2pt
            \hbox{\vrule width.#2pt height#1pt \kern5pt 
#3
                  \vrule width.#2pt}\hrule height.#2pt}}}}

\numberwithin{equation}{section}

\numberwithin{table}{section}

\begin{document}

\begin{center}

{\large\bf Anomaly resolution via decomposition}

\vspace*{0.2in}

Daniel G. Robbins$^1$, Eric Sharpe$^2$,
Thomas Vandermeulen$^1$

\begin{tabular}{cc}
{\begin{tabular}{l}
$^1$ Department of Physics\\
University at Albany\\
Albany, NY 12222 \end{tabular}} &
{\begin{tabular}{l}
$^2$ Department of Physics MC 0435\\
850 West Campus Drive\\
Virginia Tech\\
Blacksburg, VA  24061 \end{tabular}}
\end{tabular}

{\tt dgrobbins@albany.edu},
{\tt ersharpe@vt.edu},
{\tt tvandermeulen@albany.edu}

\end{center}

In this paper, we apply decomposition to orbifolds with quantum symmetries
to resolve anomalies.
Briefly, it has been argued by e.g.~Wang-Wen-Witten, Tachikawa that
an anomalous orbifold can sometimes be resolved by enlarging the orbifold
group so that the pullback of the anomaly to the larger orbifold
group is trivial.
For this
procedure to resolve the anomaly, one must specify a set of phases
in the larger orbifold, whose form is implicit in the
extension construction.  There are multiple choices of consistent 
phases,
which give rise to physically distinct resolutions.
We apply decomposition, and
find that theories with enlarged orbifold groups
are
equivalent to (disjoint unions of copies of) orbifolds by 
nonanomalous subgroups of the
original orbifold group.
In effect, decomposition implies that
enlarging the orbifold group is equivalent to making it
smaller.
We provide a general conjecture for such descriptions, which we check
in a number of examples.

\begin{flushleft}
July 2021
\end{flushleft}

\newpage

\tableofcontents

\newpage

\section{Introduction}

This paper is devoted to a study of anomalous orbifolds and their
resolutions, a subject that has been of renewed interest,
see 
e.g.~\cite{Wang:2017loc,Bhardwaj:2017xup,Tachikawa:2017gyf,Chang:2018iay,Robbins:2019zdb,Robbins:2019ayj,yujitasi2019,Robbins:2021lry,Robbins:2021ylj}.
In particular, this paper utilizes decomposition 
(which relates e.g.~two-dimensional theories with one-form symmetries
to disjoint unions)
and quantum symmetries
to make certain families of anomaly resolutions explicit,
following up our previous work
\cite{Robbins:2021ylj,Robbins:2021lry,rsv1}.

Briefly, it was argued in \cite{Tachikawa:2017gyf,Chang:2018iay} 
that given an anomalous
orbifold $[X/G]$, with anomaly $\alpha \in H^3(G,U(1))$, one way to
resolve the anomaly is to find an extension
\begin{equation}
1 \: \longrightarrow \: K \: \stackrel{\iota}{\longrightarrow} \: \Gamma \: 
\stackrel{\pi}{\longrightarrow}
\: G \: \longrightarrow \: 1,
\end{equation}
chosen so that the anomaly $\alpha \in H^3(G,U(1))$ is in the image of
some
$B \in H^1(G,H^1(K,U(1)))$, under the differential
\begin{equation}
d_2: \: H^1(G,H^1(K,U(1))) \: \longrightarrow \:
H^3(G,U(1))
\end{equation}
of the Lyndon-Hochschild-Serre spectral sequence.
(For simplicity, throughout this paper, we restrict to central extensions.)
For such an extension, the pullback $\pi^* \alpha$ is trivial in 
$H^3(\Gamma,U(1))$, and the resulting orbifold $[X/\Gamma]$ is expected
to be anomaly-free.  

Construction of suitable extensions $\Gamma$ has been discussed
elsewhere, see e.g.~\cite{Wang:2017loc,Tachikawa:2017gyf,Robbins:2019zdb,Robbins:2019ayj,Robbins:2021lry}.

However, 
as also noted in e.g.~\cite[section 2]{Tachikawa:2017gyf}, to physically 
define the $\Gamma$ orbifold, we need to specify more than just
$\Gamma$ itself, we must also specify the action of $\Gamma$.  The
$\Gamma$ action is partially specified by saying that $K$ acts
trivially on $X$; however, as we shall see explicitly in examples,
that does not uniquely specify the action, and in any event,
does not suffice to make $[X/\Gamma]$ anomaly-free in general.

First, to uniquely specify the $\Gamma$ orbifold, we interpret
$B \in H^1(G,H^1(K,U(1)))$, the same
mathematical quantity that was chosen to be in the preimage of 
the anomaly $\alpha$, 
as a 
physical quantum symmetry in the sense of \cite{rsv1}.
(The physical relevance of that mathematical quantity was also
discussed, from a slightly different perspective, in
\cite[section 2]{Tachikawa:2017gyf}.)

Second,
we apply decomposition to
explain the crucial role the quantum symmetry $B$
plays in 
resolving the anomaly.  Briefly, in this context \cite{Robbins:2021lry,Robbins:2021ylj,rsv1},
decomposition relates orbifolds with quantum symmetries to disjoint
unions of orbifolds by smaller groups.
We will see explicitly
that for $B$ such that $d_2 B = \alpha$,
if we interpret $B$ as defining a quantum symmetry,
then the resulting orbifold $[X/\Gamma]_B$ is indeed anomaly-free, and in fact
by virtue of decomposition
is equivalent to (a disjoint union of) orbifolds by anomaly-free
subgroups of $G$, subgroups such that the restriction of $\alpha$ is
trivial.  (If $\Gamma$ is a `minimal' choice of extension, then
one gets a single orbifold, but if $\Gamma$ is larger than needed, in
some sense, one gets a disjoint union of several orbifolds.)

In effect, this means that resolving an anomalous $G$ orbifold by
an orbifold by an extension $\Gamma$ is physically equivalent to 
replacing the $G$ orbifold with (copies of) an orbifold by a 
nonanomalous subgroup of $G$.  (Related observations concerning module
categories have been made in the mathematics literature, see 
e.g.~\cite[example 9.7.2]{egno}.)  Put another way, in a certain sense,
making the orbifold group larger is equivalent to making it smaller,
a duality somewhat reminiscent of T-duality on a circle, with the
anomalous orbifold playing a role analogous to the self-dual radius.  

Furthermore, which resolution is obtained depends upon the choice of 
$B$.  For a fixed anomaly $\alpha$, for a fixed extension $\Gamma$,
there may be multiple choices of quantum symmetry $B$ such that
$d_2 B = \alpha$ (not to mention choices of discrete torsion),
each of which can yield a physically distinct, nonanomalous, theory.

We also emphasize merely working with a resolution $\Gamma$
does not suffice to resolve the anomaly:
if the $G$ orbifold is anomalous, then we will see explicitly that 
taking vanishing quantum symmetry
in the $\Gamma$ orbifold yields
anomalous results, regardless of $\Gamma$.  
To resolve the anomaly, one must not only take an extension
$\Gamma$ but in addition one must pick a nonzero 
quantum symmetry $B$, such that $d_2 B$ coincides with the anomaly $\alpha$.

We begin in section~\ref{sect:rev} by reviewing quantum symmetries and
decomposition.  Specifically, in this paper we crucially use a 
generalization of ordinary quantum symmetries discussed in
\cite{Robbins:2021lry,rsv1} to make sense of the anomaly-resolution
procedure.  We also review decomposition, which relates e.g.~orbifolds
with trivially-acting subgroups to disjoint unions of copies of
orbifolds by effectively-acting groups.

In section~\ref{sect:application} we apply
decomposition to simplify the anomaly resolution procedure described
earlier.  Briefly, decomposition makes it clear that the effect of
enlarging the orbifold group and turning on a quantum symmetry is
equivalent to working with (disjoint unions of) orbifolds by
nonanomalous subgroups.
We check that prediction explicitly in section~\ref{sect:exs} in a number
of examples.

We close in appendix~\ref{app:deg3-gpcohom} with a collection of
some pertinent
results on degree-three group cohomology, as arises in describing
anomalies.

We reiterate that
throughout this paper,
we assume that $\Gamma$ is a central extension of $G$ by $K$.

Finally, a remark on nomenclature.  Across our several papers on
decomposition and quantum symmetries, we have unfortunately mixed
additive and multiplicative notations.  For example, a trivial
quantum symmetry is sometimes written additively, as $B=0$,
and sometimes multiplicatively, as $B(g) = 1$ for all $g \in G$.

\section{Review of quantum symmetries and decomposition}
\label{sect:rev}

Let us quickly review quantum symmetries in orbifolds and properties
of the resulting quantum field theories, as described in much greater
detail in \cite{rsv1}.

Quantum symmetries (as we use the term in \cite{rsv1})
arise in orbifolds in which a subgroup of the
orbifold group acts trivially.  Consider an orbifold
$[X/\Gamma]$ where $\Gamma$ is a central extension of $G$ by
$K \subset \Gamma$ in which $K$ acts trivially on $X$:
\begin{equation} 
1 \: \longrightarrow \: K \: \stackrel{\iota}{\longrightarrow} \: \Gamma \: 
\stackrel{\pi}{\longrightarrow} \:
G \: \longrightarrow \: 1.
\end{equation}
Typically in this paper, $[X/G]$ will be an anomalous orbifold,
with anomaly $\alpha \in H^3(G,U(1))$, and the extension above will be
chosen\footnote{
Details of how the resolutions are chosen are described in
e.g.~\cite[section 2.7]{Tachikawa:2017gyf},
\cite[section 5.1]{Wang:2017loc}, and more efficient versions given in
\cite{Robbins:2019zdb,Robbins:2019ayj,Robbins:2021lry}.
Our focus in this paper will be on understanding the physics of the
resolutions, not the resolutions per se.
} so as to resolve the anomaly, in part.

To resolve the orbifold, one must pick a quantum symmetry, which 
for $K$ central is a bihomomorphism 
\begin{equation}
B: \: G \times K \: \longrightarrow \: U(1)
\end{equation}
defining phases acquired by $G$-twist fields under the
action of $K$ (see also \cite[section 2]{Tachikawa:2017gyf}).  
In terms of genus-one partition functions, this means
\begin{equation}  \label{eq:quantsymm}
{\scriptstyle g z} \square_h \: = \: B(\pi(h), z) \left(
{\scriptstyle g} \square_h \right),
\: \: \:
{\scriptstyle g} \square_{hz} \: = \: B( \pi(g), z)^{-1} \left(
{\scriptstyle g} \square_h \right),
\end{equation}
for $z \in K$ and $g, h \in \Gamma$ a commuting pair.
The resulting possible values of the quantum symmetry $B$ are classified by
elements of
\begin{equation}
H^1(G, H^1(K,U(1))).
\end{equation}

Sometimes these quantum symmetries can be constructed from discrete torsion.
The relationship between quantum symmetries, discrete torsion, and
anomalies is encoded in the exact sequence
\begin{equation}  \label{eq:shortexact1}
\left( {\rm Ker}\, i^* \subset H^2(\Gamma,U(1))
\right) \: \stackrel{\beta} \longrightarrow \:
H^1(G, H^1(K,U(1))) \: \stackrel{d_2}{\longrightarrow} \:
H^3(G,U(1)),
\end{equation}
where $\iota: K \rightarrow \Gamma$ is inclusion and
in the sequence above,
\begin{equation}
\iota^*: \: H^2(\Gamma,U(1)) \: \longrightarrow \:
H^2(K,U(1)).
\end{equation}
(This is part of a seven-term exact sequence \cite{hochschild},
which was discussed in greater detail in 
our previous papers
\cite{Robbins:2021ylj,Robbins:2021lry}.)
Briefly, for discrete torsion $\omega \in {\rm Ker}\, \iota^* \subset
H^2(\Gamma,U(1))$,
its image as a quantum symmetry is\footnote{
The fact that $1/\beta$ appears rather than $\beta$ itself is a 
consequence of definitions.
}
\begin{equation}
\frac{1}{\beta(\omega)(\overline{g})} \: = \: 
\frac{ \omega(z,s(\overline{g})) }{ \omega(s(\overline{g}),z) },
\end{equation}
for $z \in K$, $\overline{g} \in G$, and $s: G \rightarrow \Gamma$
any section.  (Independence from the choice of section is ultimately due to
the fact that $\omega \in {\rm Ker}\, \iota^*$.)
Similarly, the map $d_2$, which coincides with a differential in the
Lyndon-Hochschild-Serre spectral sequence, is
\begin{equation}
\label{eq:d2BFormula}
(d_2 B)(\overline{g}_1, \overline{g}_2, \overline{g}_3) \: = \:
B\left( \overline{g}_1, s_2 s_3 s_{23}^{-1} \right),
\end{equation}
where $s_i = s(\overline{g}_i)$ and $s: G \rightarrow \Gamma$ is any
section.  (Note that the quantity $s_2 s_3 s_{23}^{-1}$ is the extension
class of the extension $\Gamma$ of $G$ by $K$.)

In this language, we see that discrete torsion (whose restriction to $K$
is trivial) defines a quantum symmetry, but the quantum symmetries we
need, with nontrivial images under $d_2$, are not produced by discrete
torsion.  The extension $\Gamma$ is chosen so that there exists a $B$
for which $d_2 B$ is the anomaly $\alpha \in H^3(G,U(1))$.
Furthermore, the reader should note that the anomaly does not uniquely
determine $B$ -- for example, it can be shifted by (the image of)
discrete torsion.  The physics of the resulting resolution will depend
upon the choice of $B$, as we shall see, not just the choice of
$\alpha = d_2 B$.

Now, orbifolds in which subgroups of the orbifold group act
trivially are equivalent to disjoint unions of orbifolds by
smaller groups, known as `universes.'  This is known as decomposition
(see  
e.g.~\cite{Hellerman:2006zs,Sharpe:2019ddn,Anderson:2013sia,Robbins:2021ylj}),
and has analogues in more general gauge theories with higher-form
symmetries and various generalizations,
as has been discussed in 
e.g.~\cite{Sharpe:2014tca,Gu:2018fpm,Gu:2020ivl,Tanizaki:2019rbk,Cherman:2020cvw,Eager:2020rra,Komargodski:2020mxz}.

In the current circumstances, for an orbifold $[X/\Gamma]$
with a quantum symmetry $B$ and discrete torsion $\omega \in
H^2(\Gamma,U(1))$, the pertinent version of decomposition
was described in \cite{rsv1}.  We will not do detailed decomposition
computations in this paper, instead citing results described elsewhere,
but for completeness, we outline the pertinent results here, which
are for the case $\iota^* \omega = 0$.
\begin{enumerate}
\item Suppose that $\iota^* \omega = 0$ and $\beta(\omega)$ is nontrivial,
then
\begin{equation}
{\rm QFT}\left( [X/\Gamma]_{B,\omega} \right) \: = \:
{\rm QFT}\left( \left[ \frac{X \times \widehat{{\rm Coker}\, (B/ \beta(\omega))} }{
{\rm Ker}\, (B/ \beta(\omega))} \right]_{\hat{\omega}_0} \right).
\end{equation}
\item Suppose that $\iota^* \omega = 0$ and
$\omega = \pi^* \overline{\omega}$ for $\overline{\omega} \in
H^2(G,U(1))$.  Then,
\begin{equation}
{\rm QFT}\left( [X/\Gamma]_{B,\omega} \right) \: = \:
{\rm QFT}\left( \left[ \frac{X \times \widehat{{\rm Coker}\, B} }{
{\rm Ker}\, B} \right]_{\overline{\omega} + \hat{\omega}_0} \right).
\end{equation}
\end{enumerate}
In each case, $\hat{\omega}_0$ denotes discrete torsion on components,
and is discussed in \cite{rsv1}.

We will use decomposition to simplify $\Gamma$ orbifolds with quantum
symmetries, and will see in examples that if $B$ resolves the anomaly
(meaning that $d_2 B = \alpha$), then the result of decomposition
will be manifestly anomaly-free orbifolds, involving anomaly-free
subgroups of $G$.

\section{Application of decomposition to anomalies}
\label{sect:application}

So far we have discussed $\Gamma$ orbifolds, central extensions of
$G$ orbifolds, with quantum symmetries defined by $B \in
H^1(G,H^1(K,U(1)))$, and described a general conjecture relating those
orbifolds to simpler orbifolds by subgroups of $G$, generalizing
decomposition
\cite{Hellerman:2006zs,Sharpe:2019ddn,Tanizaki:2019rbk,Robbins:2021ylj}.

In this section we describe the application to curing anomalies.

One way to cure an anomaly $\alpha \in H^3(G,U(1))$ in a $G$ orbifold
is to extend $G$ to a larger (finite) group $\Gamma$
with $B \in H^1(G, H^1(K,U(1)))$ such that
$d_2(B) = \alpha$.  The $B$ appearing in the mathematics
defines, physically, a quantum symmetry, as we have discussed.

Then, a good intuition for the general claim of the previous
subsection,
which we will confirm in examples,
is that $B$ reduces $G$ to a subgroup that does not participate
in the anomaly, and if $K$ is larger than needed to resolve the anomaly,
then one gets multiple copies, in the spirit of decomposition
\cite{Hellerman:2006zs,Sharpe:2019ddn,Tanizaki:2019rbk,Robbins:2021ylj}.
In particular,
from (\ref{eq:d2BFormula}), 
we have immediately that the restriction of $\alpha$ to
Ker $B$ is trivial,
\begin{equation}
\alpha |_{{\rm Ker}\, B} \: = \: 0 \: \in \: H^3( {\rm Ker}\, B, U(1)),
\end{equation}
since $\alpha = d_2(B)$ and the restriction of $[B]$ to Ker $B$ vanishes.
This guarantees that the orbifold by Ker $B \subset G$ is nonanomalous.

Now, if the $\Gamma$ orbifold is given discrete torsion $\omega$,
and if $\iota^* \omega = 0$ and $\beta(\omega) \neq 0$,
then as discussed in the previous subsection and also
\cite{Robbins:2021ylj},
$\beta(\omega)$ is an element of $H^1(G,H^1(K,U(1)))$, and so contributes
to the quantum symmetry.
However, $\beta(\omega)$ cannot change the anomaly cancellation,
as $d_2 \circ \beta = 0$.
We can see that using
the seven-term
exact sequence \cite[section I.6]{neukirch}, \cite{hochschild}
that can be derived from the Lyndon-Hochschild-Serre spectral sequence
\begin{eqnarray}
\lefteqn{
0 \: \longrightarrow \: H^1(G, U(1)) \: \stackrel{\pi^*}{\longrightarrow} \:
H^1(\Gamma, U(1)) \: \stackrel{\iota^*}{\longrightarrow} \:
H^1(K, U(1)) \: \stackrel{d_2}{\longrightarrow} \:
H^2(G, U(1))
} \nonumber \\
& &
 \: \stackrel{\pi^*}{\longrightarrow} \:
{\rm Ker}\left( \iota^*  \right)
\: \stackrel{\beta}{\longrightarrow} \:
H^1( G, H^1(K,U(1)))
\: \stackrel{d_2}{\longrightarrow} \:
H^3(G, U(1))
\end{eqnarray}
If a given element of $H^1(G, H^1(K,U(1)))$ is in the image of
$\beta$, the image of discrete torsion (whose restriction to $K$ is
trivial), then its image in $H^3(G,U(1))$ under $d_2$ must vanish.
Thus, the $B$'s that arise when trivializing anomalies are in some
sense orthogonal to those arising from discrete torsion in the
considerations of \cite{Robbins:2021ylj}.

The idea of resolving an anomalous orbifold by instead orbifolding
by a subgroup (and possibly turning on discrete torsion)
has also appeared in e.g.~\cite[section 5.3]{Bhardwaj:2017xup}.
Here we have arrived at the same result, by gauging an extension of the
anomalous group.  In some sense, we see that those two approaches to
anomalous orbifolds are equivalent to one another.
(See also \cite[example 9.7.2]{egno} for a related discussion of
module categories in the mathematics literature.)

\section{Examples}
\label{sect:exs}

\subsection{Anomalous cyclic groups extended to larger cyclic groups}

It was argued in \cite{Robbins:2019zdb},
\cite[section 3]{Robbins:2021lry}
that an anomalous ${\mathbb Z}_N$ orbifold with 
anomaly $\alpha \in H^3({\mathbb Z}_N, U(1)) = {\mathbb Z}_N$ of
order $k$ (meaning $\alpha^k = 1$) can be trivialized by an extension to
${\mathbb Z}_{kN}$, so that $\pi^* \alpha$ is trivial for $\pi:
{\mathbb Z}_{kN} \rightarrow {\mathbb Z}_N$.  Now, to actually implement
that anomaly-free resolution, one must pick quantum symmetries,
and we shall do exactly that in this section, enumerating quantum symmetries
and explaining their effects, explicitly
demonstrating that any quantum symmetry $B$
such that $d_2 B = \alpha$ resolves the anomaly in an extension to
${\mathbb Z}_{kN}$.

\subsubsection{Anomalous ${\mathbb Z}_2$ extended to ${\mathbb Z}_{2k}$}

In this section we will add anomalies to examples studied
in \cite[section 4.1]{rsv1}.  Consider an anomalous orbifold
$[X/{\mathbb Z}_2]$, with anomaly $\alpha \in H^3({\mathbb Z}_2,U(1))
= {\mathbb Z}_2$.  We resolve the anomaly by extending the
orbifold group by $K = {\mathbb Z}_k$ to $\Gamma = {\mathbb Z}_{2k}$:
\begin{equation}
1 \: \longrightarrow \: {\mathbb Z}_{k} \: \longrightarrow \:
{\mathbb Z}_{2k} \: \longrightarrow \: {\mathbb Z}_2 \:
\longrightarrow \: 1,
\end{equation}
and by turning on a quantum symmetry $B$ such that $d_2 B = \alpha$.
To be nontrivial, we assume that $k$ is even.
The case $k=2$ corresponds to a minimal resolution.

Let us also briefly comment on how this extension trivializes the
pullback of anomaly in the case $k$ is even.  Let $\pi: {\mathbb Z}_{2k}
\rightarrow {\mathbb Z}_2$ denote the projection, and note that since
$\alpha$ is of order two, $\pi^* \alpha$ must be either $1$ or $g^k$,
where $g$ generates ${\mathbb Z}_{2k}$.  One can compute a coboundary-invariant
phase, as in \cite[eq. (3.2)]{Robbins:2021lry},
to show that 
\begin{equation}
\prod_{i=1}^{2k-1} \pi^* \alpha (-1, i, 1) \: = \:
\prod_{i=1}^{2k-1} (-)^i \: = \: (-)^k,
\end{equation}
hence if $k$ is odd, then $\pi^* \alpha$ can not be trivial.
It remains to show that in the case $k$ is even,
$\pi^* \alpha = 1$.  To do so, note that $\pi$
acts as reduction mod $2$, and factors
through ${\mathbb Z}_4$ when $k$ is even:
\begin{equation}
\pi(n) \: = \: a( b(n) ),
\end{equation}
where $b: {\mathbb Z}_{2k} \rightarrow {\mathbb Z}_4$ is reduction mod $4$
and $a: {\mathbb Z}_4 \rightarrow {\mathbb Z}_2$ is reduction mod $2$.
Hence, for $k$ even,
it suffices to show that $a^* \alpha$ is trivial, and a trivialization
was given in \cite[eq. (3.10)]{Robbins:2021lry}.

Now, to physically construct the orbifold resolving the anomaly,
we are instructed to turn on any quantum symmetry
$B \in H^1(G,H^1(K,U(1)))$
in the preimage of $\alpha$ in the orbifold $[X/\Gamma]$.
For this $\Gamma$, $H^2(\Gamma,U(1)) = 0$,
so $H^1(G,H^1(K,U(1)))$ injects into $H^3(G,U(1))$.
Furthermore,
\begin{eqnarray}
H^1(G, H^1(K,U(1))) & = & {\rm Hom}\left( {\mathbb Z}_2,
H^1({\mathbb Z}_{k}, U(1)) \right) \: = \: {\rm Hom}\left( 
{\mathbb Z}_2, {\mathbb Z}_{k} \right), 
\\
& = & \left\{ \begin{array}{cl}
{\mathbb Z}_2 & k \mbox{ even}, \\
0 & k \mbox{ odd}.
\end{array} \right.
\end{eqnarray}
Thus, so long as $k$ is even, there exists precisely one $B$ that trivializes
the anomaly $\alpha$.  However, if $k$ is odd, the anomaly cannot
be trivialized by extending ${\mathbb Z}_2$ by ${\mathbb Z}_k$.

Let us assume that $k$ is even, so that the anomaly can be trivialized.

We argued in \cite[section 4.1]{rsv1} that for the possible
values of $B$,
\begin{equation}  \label{eq:z2k:predict}
{\rm QFT}\left( [X/{\mathbb Z}_{2k} ]_B \right) \: = \:
\left\{ \begin{array}{cl}
{\rm QFT}\left( \coprod_{k} [X/{\mathbb Z}_2] \right) & B = 0, 
\\
{\rm QFT}\left( \coprod_{k/2} X \right) & B \neq 0.
\end{array} \right.
\end{equation}

In this case, since $\alpha$ is nontrivial,
the $B$ that trivializes $\alpha$
is also the nontrivial element.
If instead we pick the trivial $B = 0$, 
then the resulting QFT is anomalous, as it is merely copies of
the anomalous orbifold $[X/{\mathbb Z}_2]$.  (Furthermore,
since all twisted sectors appear, there is no chance of cancelling
the anomaly merely by removing certain offending modular orbits from
the partition function.)  

On the other hand, 
for the $B$ that trivializes the anomaly, we see that the quantum field
theory is well-defined:  a sum over copies of $X$, not an anomalous orbifold
of $X$.

Thus, we see explicitly that the combination of working with a larger
orbifold group and turning on a quantum symmetry resolves the anomaly.
In the case of the minimal resolution (for which $k=2$), this resolution
is equivalent to replacing the original anomalous orbifold ($[X/{\mathbb Z}_2]$)
by an orbifold by a nonanomalous subgroup.  Since there are no nontrivial
subgroups, that means replacing $[X/{\mathbb Z}_2]$ by $X$ itself.
For nonminimal resolutions (for which $k > 2$), one simply gets copies.

\subsubsection{Anomalous ${\mathbb Z}_3$ extended to ${\mathbb Z}_9$}

In this section we start with an anomalous $G = {\mathbb Z}_3$ orbifold,
and remove the anomaly $\alpha \in H^3(G,U(1))$
by extending $G$ by $K = {\mathbb Z}_3$ to
$\Gamma = {\mathbb Z}_9$:
\begin{equation}
1 \: \longrightarrow \: {\mathbb Z}_3 \: \longrightarrow \:
{\mathbb Z}_9 \: \longrightarrow \: {\mathbb Z}_3 \: \longrightarrow \: 1,
\end{equation}
and of course turning on a suitable quantum symmetry $B$.

In this case,
\begin{equation}
B \: \in \: {\rm Hom}(G, H^1( K, U(1) ) \: = \: {\mathbb Z}_3.
\end{equation}
In \cite[section 4.1.3]{rsv1}, we computed that
\begin{equation}
{\rm QFT}\left( [X/\Gamma]_B \right) \: = \:
\left\{ \begin{array}{cl}
{\rm QFT}\left( \coprod_3 [X/{\mathbb Z}_3] \right) & B = 0, \\
{\rm QFT}\left( X \right) & B \mbox{ nontrivial}.
\end{array} \right.
\end{equation}

The anomaly is trivialized by any $B \in H^1(G, H^1(K,U(1)))$ 
such that $d_2 B = \alpha$.
In this case, $H^2(\Gamma,U(1)) = 0$, so
$H^1(G,H^1(K,U(1)))$ injects into $H^3(G,U(1))$.
In this case, since $\alpha$ is nontrivial, $B$ is also nontrivial
(but since decomposition gives the same answer for both nontrivial
values of $B$, we do not need to track which nontrivial element 
$B$ corresponds to). 

For $B = 0$, the anomaly should not be trivialized, and indeed the
QFT is just copies of the original anomalous orbifold.
For $B$ nontrivial, the QFT is well-defined, simply $X$ itself.

\subsection{Anomalous ${\mathbb Z}_2 \times {\mathbb Z}_2$ extended to
${\mathbb Z}_2 \times {\mathbb Z}_4$}

In this section, we start with an anomalous
$G = {\mathbb Z}_2 \times {\mathbb Z}_2$ orbifold, and remove
the anomaly $\alpha \in H^3(G,U(1))$ by extending by 
$K = {\mathbb Z}_2$ to $\Gamma = {\mathbb Z}_2 \times {\mathbb Z}_4$:
\begin{equation}
1 \: \longrightarrow \: {\mathbb Z}_2 \: \longrightarrow \:
{\mathbb Z}_2 \times {\mathbb Z}_4 \: \longrightarrow \:
{\mathbb Z}_2 \times {\mathbb Z}_2 \: \longrightarrow \: 1,
\end{equation}
and by turning on a suitable quantum symmetry $B$.
In this case, one can also turn on ordinary discrete torsion
$\omega \in H^2(\Gamma,U(1))$, so we have several choices we can make
to resolve the orbifold.

Write the elements of ${\mathbb Z}_2 \times {\mathbb Z}_2 = 
\langle \overline{a}, \overline{b} \rangle$.
Using the fact that
\begin{equation}
H^1(G, H^1(K,U(1))) \: = \: {\mathbb Z}_2 \times {\mathbb Z}_2,
\end{equation}
the possible values of
$B$ are
characterized by their values on $\overline{a}$, $\overline{b}$.
Furthermore, since $H^2({\mathbb Z}_2 \times {\mathbb Z}_4,U(1)) =
{\mathbb Z}_2$ (see e.g.~\cite[section D.2]{Robbins:2021ylj}), we can
also turn in discrete torsion in this theory.

This example was computed in \cite[section 4.2]{rsv1}, where it was
shown that the quantum field theory of
$[X/{\mathbb Z}_2 \times {\mathbb Z}_4]_{B,\omega}$ takes the values listed
in table~\ref{table:ex:z2z4:z2z2}.

\begin{table}[h]
\begin{center}
\begin{tabular}{c|c|cc}
$B(\overline{a})$ & $B(\overline{b})$ & Without d.t. & With d.t. \\ \hline
$+1$ & $+1$ & $\coprod_2 [X/{\mathbb Z}_2 \times {\mathbb Z}_2]$ &
$\coprod_2 [X/{\mathbb Z}_2 \times {\mathbb Z}_2]_{\rm dt}$ 
\\
$-1$ & $+1$ & $[X/{\mathbb Z}_2 = \langle \overline{b} \rangle]$ &
$[X/{\mathbb Z}_2 = \langle \overline{b} \rangle]$ 
\\
$+1$ & $-1$ & $[X/{\mathbb Z}_2 = \langle \overline{a} \rangle]$ &
$[X/{\mathbb Z}_2 = \langle \overline{a} \rangle]$
\\
$-1$ & $-1$ & $[X/{\mathbb Z}_2 = \langle \overline{a} \overline{b} \rangle]$ &
$[X/{\mathbb Z}_2 = \langle \overline{a} \overline{b} \rangle]$
\end{tabular}
\caption{Summary of decomposition results for $[X/{\mathbb Z}_2 \times {\mathbb Z}_4]_{B,\omega}$
for various values of $B$, $\omega$, from \cite[section 4.2]{rsv1}.
For nontrivial $B$, adding discrete torsion has no effect in these cases.
\label{table:ex:z2z4:z2z2}
}
\end{center}
\end{table}

To understand how quantum symmetries can resolve anomalies in this
case, we next compute $d_2 B$.  We will do so manually.
In principle,
\begin{equation}
(d_2 B)(\overline{g}_1, \overline{g}_2, \overline{g}_3)
\: = \: B( \overline{g}_1, e(\overline{g}_2, \overline{g}_3) ),
\end{equation}
for $\overline{g}_i \in G$, and where $e$ denotes the extension class
of $\Gamma$.
Recall that for a section $s: G \rightarrow \Gamma$,
the extension class $e$ is given explicitly as a cocycle\footnote{
This cochain is coclosed so long as $K$ is central, which we assume
throughout this paper.
}
\begin{equation}
e(\overline{g}_1,\overline{g}_2) \: = \:
s_1 s_2 s_{12}^{-1},
\end{equation}
where $s_i = s(\overline{g}_i)$. 
For this case, $\Gamma = {\mathbb Z}_2 \times {\mathbb Z}_4$, where
$\Gamma = \langle a, b \rangle$, $a^2 = 1 = b^4$.  In these conventions,
$K = {\mathbb Z}_2 = \langle b^2 \rangle$.  The projection $\pi$
maps $a \mapsto \overline{a}$, $b \mapsto \overline{b}$, and we pick 
the section $s$ given by
\begin{equation}
s(1) \: = \: 1, \: \: \:
s(\overline{a}) \: = \: a, \: \: \:
s(\overline{b}) \: = \: b, \: \: \:
s(\overline{a} \overline{b}) \: = \: ab.
\end{equation}
Then, we compute explicitly that the values of $e(\overline{g}_1,\overline{g}_2)$
are given as in the table below:
\begin{center}
\begin{tabular}{c|cccc}
$e$ & $1$ & $\overline{a}$ & $\overline{b}$ & $\overline{a} \overline{b}$
\\ \hline
$1$ & $1$ & $1$ & $1$ & $1$ \\
$\overline{a}$ & $1$ & $1$ & $1$ & $1$ \\
$\overline{b}$ & $1$ & $1$ & $b^2$ & $b^2$ \\
$\overline{a} \overline{b}$ & $1$ & $1$ & $b^2$ & $b^2$
\end{tabular}
\end{center}

Next, we compute $d_2 B$ for various choices of $B$.
\begin{enumerate}
\item In the trivial case $B(\overline{a}) = +1 = B(\overline{b})$,
we find that $d_2 B( \overline{g}_1, \overline{g}_2, \overline{g}_3) = 1$,
and so the cohomology class of $d_2 B$ is trivial.
\item Consider the case $B(\overline{a}) = -1$, $B(\overline{b}) = +1$.
In this case it is straightforward to compute
\begin{equation}
(d_2 B)(\overline{a},\overline{b},\overline{b}) \: = \: -1
\: = \: (d_2 B)(\overline{a},\overline{b},\overline{a}\overline{b})
\: = \: (d_2 B)(\overline{a},\overline{a}\overline{b}, \overline{b})
\: = \: (d_2 B)(\overline{a},\overline{a}\overline{b},\overline{a}\overline{b}),
\end{equation}
\begin{equation}
(d_2 B)(\overline{a}\overline{b},\overline{b},\overline{b}) \: = \: -1
\: = \: (d_2 B)(\overline{a}\overline{b},\overline{b},\overline{a}\overline{b})
\: = \: (d_2 B)(\overline{a}\overline{b},\overline{a}\overline{b}, \overline{b})
\: = \: (d_2 B)(\overline{a}\overline{b},\overline{a}\overline{b},\overline{a}\overline{b}),
\end{equation}
with all other entries $+1$.
From the invariants~(\ref{eq:invt1})-(\ref{eq:invt6}),
we find that the cohomology class of $d_2 B \in H^3({\mathbb Z}_2 \times
{\mathbb Z}_2,U(1))$ is characterized by 
$\epsilon_{\overline{a}\overline{b}} = -1$,
$\epsilon_{\overline{a}} = +1 = \epsilon_{\overline{b}}$.
Thus, for this $B$, $d_2 B$ is nontrivial in cohomology, and can be
used to resolve an anomalous $\langle \overline{a} \overline{b} \rangle
\subset {\mathbb Z}_2 \times {\mathbb Z}_2$.
\item Next, consider the case $B(\overline{a}) = +1$,
$B(\overline{b}) = -1$.
In this case it is straightforward to compute
\begin{equation}
(d_2 B)(\overline{b},\overline{b},\overline{b}) \: = \: -1
\: = \: (d_2 B)(\overline{b},\overline{b},\overline{a}\overline{b})
\: = \: (d_2 B)(\overline{b},\overline{a}\overline{b}, \overline{b})
\: = \: (d_2 B)(\overline{b},\overline{a}\overline{b},\overline{a}\overline{b}),
\end{equation}
\begin{equation}
(d_2 B)(\overline{a}\overline{b},\overline{b},\overline{b}) \: = \: -1
\: = \: (d_2 B)(\overline{a}\overline{b},\overline{b},\overline{a}\overline{b})
\: = \: (d_2 B)(\overline{a}\overline{b},\overline{a}\overline{b}, \overline{b})
\: = \: (d_2 B)(\overline{a}\overline{b},\overline{a}\overline{b},\overline{a}\overline{b}),
\end{equation}
with all other entries $+1$.
From the invariants~(\ref{eq:invt1})-(\ref{eq:invt6}),
we find that the cohomology class of $d_2 B \in H^3({\mathbb Z}_2 \times
{\mathbb Z}_2,U(1))$ is characterized by 
$\epsilon_{\overline{b}} = -1 = \epsilon_{\overline{a}\overline{b}}$,
$\epsilon_{\overline{a}} = +1$.
Thus, for this $B$, $d_2 B$ is nontrivial in cohomology, and can be
used to resolve an anomalous $\langle \overline{b} \rangle,
\langle \overline{a} \overline{b} \rangle \subset
{\mathbb Z}_2 \times {\mathbb Z}_2$.
\item Next, consider the case
$B(\overline{a}) = -1 = B(\overline{b})$.
In this case it is straightforward to compute
\begin{equation}
(d_2 B)(\overline{a},\overline{b},\overline{b}) \: = \: -1
\: = \: (d_2 B)(\overline{a},\overline{b},\overline{a}\overline{b})
\: = \: (d_2 B)(\overline{a},\overline{a}\overline{b}, \overline{b})
\: = \: (d_2 B)(\overline{a},\overline{a}\overline{b},\overline{a}\overline{b}),
\end{equation}
\begin{equation}
(d_2 B)(\overline{b},\overline{b},\overline{b}) \: = \: -1
\: = \: (d_2 B)(\overline{b},\overline{b},\overline{a}\overline{b})
\: = \: (d_2 B)(\overline{b},\overline{a}\overline{b}, \overline{b})
\: = \: (d_2 B)(\overline{b},\overline{a}\overline{b},\overline{a}\overline{b}),
\end{equation}
with all other entries $+1$.
From the invariants~(\ref{eq:invt1})-(\ref{eq:invt6}), 
we find that the cohomology class of $d_2 B \in H^3({\mathbb Z}_2 \times
{\mathbb Z}_2,U(1))$ is characterized by
$\epsilon_{\overline{b}} = -1$,
$\epsilon_{\overline{a}} = +1 = \epsilon_{\overline{a} \overline{b}}$.
Thus, for this $B$, $d_2 B$ is nontrivial in cohomology, and can be
used to resolve an anomalous $\langle \overline{b} \rangle
\subset {\mathbb Z}_2 \times {\mathbb Z}_2$.
\end{enumerate}

Now, let us compare to the results of decomposition listed
in table~\ref{table:ex:z2z4:z2z2}.
\begin{enumerate}
\item For the first choice of $B$, for which $B(\overline{a}) = +1 =
B(\overline{b})$, the cohomology class of $d_2 B$ is trivial, and so no 
anomaly resolution is guaranteed by the method we outlined.  Comparing to the
results of decomposition, we see that copies of the
anomalous orbifold $[X/{\mathbb Z}_2 \times {\mathbb Z}_2]$ appear,
consistent with the fact that in this case, no anomalies are expected
to be resolved.
\item In the next case, for which
$B(\overline{a}) = -1$ and $B(\overline{b}) = +1$,
an anomaly in the subgroup $\langle \overline{a} \overline{b} \rangle$
can be resolved, but the other subgroups $\langle \overline{a} \rangle$
and $\langle \overline{b} \rangle$ are assumed nonanomalous.
In this case, the physical theory for this quantum symmetry is the
orbifold $[X/\langle \overline{b} \rangle]$, which is anomaly-free.
\item In the next case, for which
$B(\overline{a}) = +1$ and $B(\overline{b}) = -1$,
an anomaly in the subgroups $\langle \overline{b} \rangle$,
$\langle \overline{a} \overline{b} \rangle$ can be resolved,
but the other subgroup $\langle \overline{a} \rangle$ is
assumed nonanomalous.  In this case, the physical theory for this
quantum symmetry is the orbifold $[X/\langle \overline{a} \rangle]$,
which is anomaly-free.
\item In the last case, for which
$B(\overline{a}) = -1 = B(\overline{b})$,
an anomaly in the subgroup $\langle \overline{b} \rangle$ can be resolved,
but the other subgroups $\langle \overline{a} \rangle$,
$\langle \overline{a} \overline{b} \rangle$ are assumed nonanomalous.
In this case, the physical theory for the quantum symmetry is the
orbifold $[X/\langle \overline{a} \overline{b} \rangle]$,
which is anomaly-free.
\end{enumerate}
In each case, the effect of turning on a quantum symmetry is to 
reduce the orbifold group to a nonanomalous subgroup
(correlated to the choice of $B$), precisely as advertised.

\subsection{Anomalous ${\mathbb Z}_2 \times {\mathbb Z}_2$ extended to
$D_4$}

In this section, we start with an anomalous $G = {\mathbb Z}_2 \times {\mathbb Z}_2$
orbifold, and remove the anomaly $\alpha \in H^3(G,U(1))$ by extending
by $K = {\mathbb Z}_2$ to $\Gamma = D_4$:
\begin{equation}
1 \: \longrightarrow \: {\mathbb Z}_2 \: \longrightarrow \:
D_4 \: \longrightarrow \: {\mathbb Z}_2 \times {\mathbb Z}_2 
\: \longrightarrow \: 1,
\end{equation}
and by turning on a suitable quantum symmetry $B$.  
In this case, one can also turn on ordinary discrete torsion
$\omega \in H^2(\Gamma,U(1))$, so we have several choices we can make
to resolve the orbifold.

Write the elements of ${\mathbb Z}_2 \times {\mathbb Z}_2 = 
\langle \overline{a}, \overline{b} \rangle$.  
Using the fact that
\begin{equation}
H^1(G, H^1(K,U(1))) \: = \: {\mathbb Z}_2 \times {\mathbb Z}_2,
\end{equation}
the possible values of $B$
are
characterized by their values on $\overline{a}, \overline{b}$.

This example was computed in \cite[section 4.3]{rsv1}, where from both
decomposition and explicit computation it was demonstrated that the
quantum field theory of $[X/\Gamma]_{B, \omega}$ took the values
listed in table~\ref{table:ex:d4:z2z2}.

\begin{table}[h]
\begin{center}
\begin{tabular}{c|c|cc}
$B(\overline{a})$ & $B(\overline{b})$ & W/o d.t. & With d.t. \\ \hline
$+1$ & $+1$ & $[X/{\mathbb Z}_2 \times {\mathbb Z}_2] \, \coprod \,
[X/{\mathbb Z}_2 \times {\mathbb Z}_2]_{\rm dt}$ &
$[X/{\mathbb Z}_2 = \langle \overline{b} \rangle]$ 
\\
$-1$ & $+1$ & $[X/{\mathbb Z}_2 = \langle \overline{b} \rangle]$ &
$[X/{\mathbb Z}_2 \times {\mathbb Z}_2] \, \coprod \,
[X/{\mathbb Z}_2 \times {\mathbb Z}_2]_{\rm dt}$
\\
$+1$ & $-1$ & $[X/{\mathbb Z}_2 = \langle \overline{a} \rangle]$ &
$[X/{\mathbb Z}_2 = \langle \overline{a} \overline{b} \rangle]$ 
\\
$-1$ & $-1$ & $[X/{\mathbb Z}_2 = \langle \overline{a} \overline{b} \rangle]$ 
&
$[X/{\mathbb Z}_2 = \langle \overline{a} \rangle]$
\end{tabular}
\caption{Summary of decomposition results for $[X/D_4]_{B,\omega}$ for various
values of $B$, $\omega$, from \cite[section 4.3]{rsv1}.
\label{table:ex:d4:z2z2} }
\end{center}
\end{table}

To understand how quantum symmetries can resolve anomalies in this
case, we next compute $d_2 B$.  We will do so manually, as we did in
the last example.
In principle,
\begin{equation}
(d_2 B)(\overline{g}_1, \overline{g}_2, \overline{g}_3)
\: = \: B( \overline{g}_1, e(\overline{g}_2, \overline{g}_3) ),
\end{equation}
for $\overline{g}_i \in G$, and where $e$ denotes the extension class
of $\Gamma$,
\begin{equation}
e(\overline{g}_1,\overline{g}_2) \: = \:
s_1 s_2 s_{12}^{-1},
\end{equation}
where $s_i = s(\overline{g}_i)$ for $s$ a section.  
For this case, $\Gamma = D_4$ and
$G = {\mathbb Z}_2 \times {\mathbb Z}_2$, write
\begin{equation}
D_4 \: = \: \{1, a, b, z, az, bz, ab, ba=abz \},
\end{equation}
where $z^2 = 1 = a^2$ and $b^2 = z$, $z$ generates the center,
$\pi(a) = \overline{a}$, and $\pi(b) = \overline{b}$.
Then, take the section to be given by
\begin{equation}
s(\overline{a}) \: = \: a, \: \: \:
s(\overline{b}) \: = \: b, \: \: \:
s(\overline{a}\overline{b}) \: = \: ab.
\end{equation}

Then, we compute explicitly that the values of $e(\overline{g}_1,\overline{g}_2)$ are given as in the table below:
\begin{center}
\begin{tabular}{c|cccc}
$e$ & $1$ & $\overline{a}$ & $\overline{b}$ & $\overline{ab}$ \\ \hline
$1$ & $1$ & $1$ & $1$ & $1$ \\
$\overline{a}$ & $1$ & $1$ & $1$ & $1$ \\
$\overline{b}$ & $1$ & $z$ & $z$ & $1$ \\
$\overline{a}\overline{b}$ & $1$ & $z$ & $z$ & $1$
\end{tabular}
\end{center}

Next, we compute $d_2 B$ for various choices of $B$.
\begin{enumerate}
\item In the trivial case $B(\overline{a}) = +1 = B(\overline{b})$,
we find that $d_2 B( \overline{g}_1, \overline{g}_2, \overline{g}_3) = 1$,
and so the cohomology class of $d_2 B$ is trivial.
\item Consider the case $B(\overline{a}) = -1$, $B(\overline{b}) = +1$.
In this case it is straightforward to compute
\begin{equation}
(d_2 B)(\overline{a}, \overline{b}, \overline{a}) \: = \: -1
\: = \: (d_2 B)( \overline{a}, \overline{b}, \overline{b})
\: = \: (d_2 B)( \overline{a}, \overline{a} \overline{b}, \overline{a})
\: = \: (d_2 B)( \overline{a}, \overline{a} \overline{b}, \overline{b}),
\end{equation}
\begin{equation}
(d_2 B)(\overline{a} \overline{b}, \overline{b}, \overline{a}) \: = \: -1
\: = \: (d_2 B)( \overline{a} \overline{b}, \overline{b}, \overline{b})
\: = \: (d_2 B)( \overline{a} \overline{b}, \overline{a} \overline{b}, \overline{a})
\: = \: (d_2 B)( \overline{a} \overline{b}, \overline{a} \overline{b}, \overline{b}),
\end{equation}
with all other entries $+1$. 
From the invariants~(\ref{eq:invt1})-(\ref{eq:invt6}),
we find that the cohomology class of $d_2 B \in H^3({\mathbb Z}_2 \times
{\mathbb Z}_2,U(1))$ is characterized by $\epsilon_{\overline{a}} = +1 = 
\epsilon_{\overline{b}} =
\epsilon_{\overline{a} \overline{b}}$, 
and so for this $B$, $d_2 B$ is trivial in cohomology.
(In fact, since this choice of $B$ is equivalent to turning on
discrete torsion, it should not be a surprise that it is in the kernel
of $d_2$.)
\item Next, consider the case $B(\overline{a}) = +1$,
$B(\overline{b}) = -1$.  In this case it is straightforward to compute
\begin{equation}
(d_2 B)(\overline{b}, \overline{b}, \overline{a}) \: = \: -1
\: = \: (d_2 B)(\overline{b}, \overline{b}, \overline{b})
\: = \: (d_2 B)(\overline{b}, \overline{a} \overline{b}, \overline{a})
\: = \: (d_2 B)(\overline{b}, \overline{a} \overline{b}, \overline{b}),
\end{equation}
\begin{equation}
(d_2 B)(\overline{a} \overline{b},  \overline{b}, \overline{a}) \: = \: -1
\: = \: (d_2 B)( \overline{a} \overline{b}, \overline{b}, \overline{b})
\: = \: (d_2 B)(\overline{a} \overline{b}, \overline{a} \overline{b}, \overline{a})
\: = \: (d_2 B)(\overline{a} \overline{b}, \overline{a} \overline{b}, \overline{b}),
\end{equation}
with all other entries $+1$.
From the invariants~(\ref{eq:invt1})-(\ref{eq:invt6}),
we find that the cohomology class of $d_2 B \in H^3({\mathbb Z}_2 \times
{\mathbb Z}_2,U(1))$ is characterized by
$\epsilon_{\overline{b}} = -1$, $\epsilon_{\overline{a}} = 
\epsilon_{\overline{a} \overline{b}} = +1$,
and so for this $B$, $d_2 B$ is nontrivial in cohomology,
and can be used to resolve anomalies in the 
$\langle \overline{b} \rangle \subset {\mathbb Z}_2 \times {\mathbb Z}_2$
subgroup.
\item Finally, consider the case $B(\overline{a}) = -1$,
$B(\overline(b)) = -1$.  In this case it is straightforward to compute
\begin{equation}
(d_2 B)(\overline{a}, \overline{b}, \overline{a}) \: = \: -1
\: = \: (d_2 B)(\overline{a}, \overline{b}, \overline{b})
\: = \: (d_2 B)(\overline{a}, \overline{a} \overline{b}, \overline{a})
\: = \: (d_2 B)(\overline{a}, \overline{a} \overline{b}, \overline{b}),
\end{equation}
\begin{equation}
(d_2 B)(\overline{b},  \overline{b}, \overline{a}) \: = \: -1
\: = \: (d_2 B)(\overline{b}, \overline{b}, \overline{b})
\: = \: (d_2 B)( \overline{b}, \overline{a} \overline{b}, \overline{a})
\: = \: (d_2 B)(\overline{b}, \overline{a} \overline{b}, \overline{b}),
\end{equation}
with all other entries $+1$.
From the invariants~(\ref{eq:invt1})-(\ref{eq:invt6}),
we find that the cohomology class of $d_2 B \in H^3({\mathbb Z}_2 \times
{\mathbb Z}_2,U(1))$ is characterized by
$\epsilon_{\overline{b}} = -1$, $\epsilon_{\overline{a}} = 
\epsilon_{\overline{a} \overline{b}} = +1$,
the same as for the previous choice of $B$,
and so for this $B$, $d_2 B$ is nontrivial in cohomology,
and again can be used to resolve anomalies in the 
$\langle \overline{b} \rangle \subset {\mathbb Z}_2 \times {\mathbb Z}_2$
subgroup.
\end{enumerate}

Now, let us compare to the results of decomposition listed
in table~\ref{table:ex:d4:z2z2}.  For the first two choices of $B$,
for which
$B(\overline{a}) = +1$, 
the cohomology class of $d_2 B$ is trivial, and so no anomaly resolution is
guaranteed by the method we outlined.  Comparing to the results of
decomposition, for these two choices of $B$, we see that for some
choices of discrete torsion, copies of the anomalous orbifold
$[X/{\mathbb Z}_2 \times {\mathbb Z}_2]$ appear, consistent with the
fact that the method described is not anticipated to resolve any anomalies.
(Curiously, for other values of discrete torsion, only the
orbifold $[X/{\mathbb Z}_2 = \langle \overline{b} \rangle]$ appears,
which would resolve an anomaly that is in the $\langle \overline{a}
\rangle$ or $\langle \overline{a} \overline{b} \rangle
 \subset {\mathbb Z}_2 \times {\mathbb Z}_2$.)

For the last two choices of $B$, for which $B(\overline{b}) = -1$,
the cohomology class of $d_2 B$ is nontrivial,
characterized by $\epsilon_{\overline{b}}
= -1$, and so we expect that these choices of $B$ can resolve an anomaly
in $\langle \overline{b} \rangle \subset {\mathbb Z}_2 \times {\mathbb Z}_2$,
so long as $\langle \overline{a} \rangle$ and $\langle \overline{a}
\overline{b} \rangle$ are anomaly-free.
This is indeed consistent with the results of decomposition in
table~\ref{table:ex:d4:z2z2}:  for both pertinent choices of $B$,
for all choices of discrete torsion, the QFTs are consistent,
in that they only involve orbifolds by anomaly-free subgroups of
${\mathbb Z}_2 \times {\mathbb Z}_2$.

Thus, to summarize, we see explicitly that our prescription works in this
example: for choices of $B$ such that $d_2 B$ matches the anomaly,
the orbifold $[X/\Gamma]_{B,\omega}$ is indeed anomaly-free, and its
precise physics is determined by choices of $B$ (with fixed image under
$d_2$) and discrete torsion $\omega$.
The reader should further note that this particular extension
($D_4$) can only be used to resolve anomalies in the
subgroup $\langle \overline{b} \rangle \subset {\mathbb Z}_2 \times
{\mathbb Z}_2$; for other anomalous subgroups, different resolutions
are required.

This example was also studied, from a different perspective,
in \cite[section 5.2.1]{Robbins:2021lry}.

\subsection{Anomalous ${\mathbb Z}_2 \times {\mathbb Z}_2$
extended to ${\mathbb H}$}

Next, we consider an example that is closely related to the previous one.
Here, we again start with an anomalous $G = {\mathbb Z}_2 \times {\mathbb Z}_2$
orbifold, and this time remove the anomaly $\alpha \in H^3(G,U(1))$ by
extending by $K = {\mathbb Z}_2$ to $\Gamma = {\mathbb H}$,
the eight-element finite group of quaternions,
\begin{equation}
1 \: \longrightarrow \: {\mathbb Z}_2 \: \longrightarrow \: 
{\mathbb H} \: \stackrel{\pi}{\longrightarrow} \: {\mathbb Z}_2 \times {\mathbb Z}_2
\: \longrightarrow \: 1,
\end{equation}
and by turning on a suitable quantum symmetry $B$.  
This extension is different from that discussed in the previous section,
and so we will see a different pattern of anomaly resolution.

As before, write the elements of ${\mathbb Z}_2 \times {\mathbb Z}_2 = 
\langle \overline{a}, \overline{b} \rangle$, so that 
using
\begin{equation}
H^1(G, H^1(K,U(1))) \: = \: {\mathbb Z}_2 \times {\mathbb Z}_2,
\end{equation}
the possible values of
$B$ are 
characterized by their values on $\overline{a}$, $\overline{b}$.
In our conventions, 
\begin{equation}
\pi(\pm i) \: = \: \overline{a}, \: \: \:
\pi(\pm j) \: = \: \overline{b}, \: \: \:
\pi(\pm k) \: = \: \overline{a} \overline{b}.
\end{equation}

Unlike the case of $D_4$, the group ${\mathbb H}$ does not admit
discrete torsion:  $H^2({\mathbb H},U(1)) = 0$.  
Applying decomposition \cite{Robbins:2021ylj}, it is straightforward
to show that QFT$( [X/{\mathbb H}]_B )$ for the various choices of $B$ are
as given in table~\ref{table:ex:h:z2z2}.

\begin{table}[h]
\begin{center}
\begin{tabular}{c|c|c}
$B(\overline{a})$ & $B(\overline{b})$ & Theory \\ \hline
$+1$ & $+1$ & $[X/{\mathbb Z}_2 \times {\mathbb Z}_2] \, \coprod \,
[X/{\mathbb Z}_2 \times {\mathbb Z}_2]_{\rm dt}$ \\
$-1$ & $+1$ & $[X/{\mathbb Z}_2 = \langle \overline{b} \rangle]$ \\
$+1$ & $-1$ & $[X/{\mathbb Z}_2 = \langle \overline{a} \rangle]$ \\
$-1$ & $-1$ & $[X/{\mathbb Z}_2 = \langle \overline{a} \overline{b} \rangle]$
\end{tabular}
\caption{Summary of decomposition results for $[X/{\mathbb H}]_B$ for various values of
$B$.  The trivial case, $B(\overline{a}) = +1 = B(\overline{b})$,
is discussed in \cite[section 5.3]{Hellerman:2006zs}.
\label{table:ex:h:z2z2}
}
\end{center}
\end{table}

To understand how quantum symmetries can resolve anomalies in this case,
we next compute $d_2 B$.  As before, in principle,
\begin{equation}
(d_2 B)(\overline{g}_1, \overline{g}_2, \overline{g}_3) \: = \:
B( \overline{g}_1, e(\overline{g}_2, \overline{g}_3) ),
\end{equation}
for $\overline{g}_i \in G$, and where $e$ denotes the extension class
of $\Gamma$.  What distinguishes this example from the previous one is
that the extension class $e$ is different.

Let us compute the extension class $e$ explicitly.  For a section $s:
G \rightarrow \Gamma$, it can be written
\begin{equation}
e(\overline{g}_1,\overline{g}_2) \: = \:
s_1 s_2 s_{12}^{-1},
\end{equation}
where $s_i = s(\overline{g}_i)$. We pick the section given by
\begin{equation}
s(1) \: = \: 1, \: \: \:
s(\overline{a}) \: = \: i, \: \: \:
s(\overline{b}) \: = \: j, \: \: \:
s(\overline{k}) \: = \: k.
\end{equation}
Then, we compute explicitly that the values of $e(\overline{g}_1, 
\overline{g}_2)$ are given as in the table below:
\begin{center}
\begin{tabular}{c|rrrr}
$e$ & $1$ & $\overline{a}$ & $\overline{b}$ & $\overline{ab}$ \\ \hline
$1$ & $1$ & $1$ & $1$ & $1$ \\
$\overline{a}$ & $1$ & $-1$ & $1$ & $-1$ \\
$\overline{b}$ & $1$ & $-1$ & $-1$ & $1$ \\
$\overline{a}\overline{b}$ & $1$ & $1$ & $-1$ & $-1$ 
\end{tabular}
\end{center}
One sees immediately that this extension class differs from that
appearing in the $D_4$ extension we considered in the previous section.

Next, we compute $d_2 B$ for various choices of $B$.
\begin{enumerate}
\item In the trivial case $B(\overline{a}) = +1 = B(\overline{b})$,
we find that $d_2 B( \overline{g}_1, \overline{g}_2, \overline{g}_3) = 1$,
and so the cohomology class of $d_2 B$ is trivial.
\item Consider the case $B(\overline{a}) = -1$, $B(\overline{b}) = +1$.
In this case it is straightforward to compute
\begin{eqnarray}
(d_2 B)(\overline{a},\overline{a},\overline{a}) \: = \: -1
& = & (d_2 B)(\overline{a},\overline{a},\overline{a}\overline{b})
\: = \: (d_2 B)(\overline{a},\overline{b},\overline{a}),
\\
& = & (d_2 B)(\overline{a},\overline{b},\overline{b})
\: = \: (d_2 B)(\overline{a},\overline{a}\overline{b},\overline{b})
\: = \: (d_2 B)(\overline{a},\overline{a}\overline{b},\overline{a}\overline{b}),
\nonumber
\\
(d_2 B)(\overline{a}\overline{b},\overline{a},\overline{a}) \: = \: -1
& = & (d_2 B)(\overline{a}\overline{b},\overline{a},\overline{a}\overline{b})
\: = \: (d_2 B)(\overline{a}\overline{b},\overline{b},\overline{a}),
\\
& = & (d_2 B)(\overline{a}\overline{b},\overline{b},\overline{b})
\: = \: (d_2 B)(\overline{a}\overline{b},\overline{a}\overline{b},\overline{b})
\: = \: (d_2 B)(\overline{a}\overline{b},\overline{a}\overline{b},\overline{a}\overline{b}), \nonumber
\end{eqnarray}
with all other entries $+1$.
From the invariants~(\ref{eq:invt1})-(\ref{eq:invt6}),
we find that the cohomology class of $d_2 B \in H^3({\mathbb Z}_2 \times
{\mathbb Z}_2,U(1))$ is characterized by 
$\epsilon_{\overline{a}} = -1 = \epsilon_{\overline{a}\overline{b}}$,
$\epsilon_{\overline{b}} = +1$,
hence for this choice of $B$, $d_2 B$ is nontrivial in cohomology,
and can be used to resolve anomalies in the
subgroups $\langle \overline{a} \rangle, \langle \overline{a} \overline{b}
\rangle \subset {\mathbb Z}_2 \times {\mathbb Z}_2$.
\item Next, consider the case $B(\overline{a}) = +1$,
$B(\overline{b}) = -1$.  In this case it is straightforward to compute
\begin{eqnarray}
(d_2 B)(\overline{b},\overline{a},\overline{a}) \: = \: -1
& = & (d_2 B)(\overline{b},\overline{a},\overline{a}\overline{b})
\: = \: (d_2 B)(\overline{b},\overline{b},\overline{a}),
\\
& = & (d_2 B)(\overline{b},\overline{b},\overline{b})
\: = \: (d_2 B)(\overline{b},\overline{a}\overline{b},\overline{b})
\: = \: (d_2 B)(\overline{b},\overline{a}\overline{b},\overline{a}\overline{b}),
\nonumber
\\
(d_2 B)(\overline{a}\overline{b},\overline{a},\overline{a}) \: = \: -1
& = & (d_2 B)(\overline{a}\overline{b},\overline{a},\overline{a}\overline{b})
\: = \: (d_2 B)(\overline{a}\overline{b},\overline{b},\overline{a}),
\\
& = & (d_2 B)(\overline{a}\overline{b},\overline{b},\overline{b})
\: = \: (d_2 B)(\overline{a}\overline{b},\overline{a}\overline{b},\overline{b})
\: = \: (d_2 B)(\overline{a}\overline{b},\overline{a}\overline{b},\overline{a}\overline{b}), \nonumber
\end{eqnarray}
with all other entries $+1$.
From the invariants~(\ref{eq:invt1})-(\ref{eq:invt6}),
we find that the cohomology class of $d_2 B \in H^3({\mathbb Z}_2 \times
{\mathbb Z}_2,U(1))$ is characterized by
$\epsilon_{\overline{b}} = -1 = \epsilon_{\overline{a}\overline{b}}$,
$\epsilon_{\overline{a}} = +1$,
hence for this choice of $B$, $d_2 B$ is nontrivial in cohomology,
and can be used to resolve anomalies in the
subgroups $\langle \overline{b} \rangle, \langle \overline{a} \overline{b}
\rangle \subset {\mathbb Z}_2 \times {\mathbb Z}_2$.
\item Finally, consider the case
$B(\overline{a}) = -1 = B(\overline{b})$.  In this case it is straightforward
to compute
\begin{eqnarray}
(d_2 B)(\overline{a},\overline{a},\overline{a}) \: = \: -1
& = & (d_2 B)(\overline{a},\overline{a},\overline{a}\overline{b})
\: = \: (d_2 B)(\overline{a},\overline{b},\overline{a}),
\\
& = & (d_2 B)(\overline{a},\overline{b},\overline{b})
\: = \: (d_2 B)(\overline{a},\overline{a}\overline{b},\overline{b})
\: = \: (d_2 B)(\overline{a},\overline{a}\overline{b},\overline{a}\overline{b}),
\nonumber
\\
(d_2 B)(\overline{b},\overline{a},\overline{a}) \: = \: -1
& = & (d_2 B)(\overline{b},\overline{a},\overline{a}\overline{b})
\: = \: (d_2 B)(\overline{b},\overline{b},\overline{a}),
\\
& = & (d_2 B)(\overline{b},\overline{b},\overline{b})
\: = \: (d_2 B)(\overline{b},\overline{a}\overline{b},\overline{b})
\: = \: (d_2 B)(\overline{b},\overline{a}\overline{b},\overline{a}\overline{b}),
\nonumber
\end{eqnarray} 
with all other entries $+1$.
From the invariants~(\ref{eq:invt1})-(\ref{eq:invt6}),
we find that the cohomology class of $d_2 B \in H^3({\mathbb Z}_2 \times
{\mathbb Z}_2,U(1))$ is characterized by
$\epsilon_{\overline{a}} = -1 = \epsilon_{\overline{b}}$,
$\epsilon_{\overline{a}\overline{b}} = +1$,
hence for this choice of $B$, $d_2 B$ is nontrivial in cohomology,
and can be used to resolve anomalies in the
subgroups $\langle \overline{a} \rangle, \langle \overline{b}
\rangle \subset {\mathbb Z}_2 \times {\mathbb Z}_2$.
\end{enumerate}

Now, let us compare to the results of decomposition listed in
table~\ref{table:ex:h:z2z2}.  
\begin{enumerate}
\item For the first choice of $B$, for which $B(\overline{a}) = +1 =
B(\overline{b})$, the cohomology class of $d_2 B$ is trivial, and so no
anomaly resolution is guaranteed by the method we outlined.
Comparing to the results of decomposition, we see that copies of
the anomalous orbifold $[X/{\mathbb Z}_2 \times {\mathbb Z}_2]$ appear,
consistent with the fact that in this case, no anomalies are expected
to be resolved.  
\item In the next case, for which $B(\overline{a}) = -1$ and
$B(\overline{b})=+1$, the subgroups $\langle \overline{a} \rangle$,
$\langle \overline{a} \overline{b} \rangle$ are (potentially)
anomalous, but 
$\langle \overline{b} \rangle$ is nonanomalous, and indeed
the physical theory for this quantum symmetry involves only an orbifold
by the nonanomalous subgroup $\langle \overline{b} \rangle$.
\item In the next case, for which $B(\overline{a}) = +1$ and
$B(\overline{b}) = -1$, the subgroups $\langle \overline{b} \rangle$,
$\langle \overline{a} \overline{b} \rangle$ are (potentially)
anomalous, but $\langle \overline{a} \rangle$ is nonanomalous,
and indeed the physical theory for this quantum symmetry involves
only an orbifold by the nonanomalous subgroup $\langle \overline{a} \rangle$.
\item In the last case, for which $B(\overline{a}) = -1 = 
B(\overline{b})$, the subgroups $\langle \overline{a} \rangle$,
$\langle \overline{b} \rangle$ are (potentially) anomalous, but
$\langle \overline{a} \overline{b} \rangle$ is nonanomalous,
and indeed the physical theory for this quantum symmetry involves
only an orbifold by the nonanomalous subgroup $\langle \overline{a} \overline{b}
\rangle$.
\end{enumerate}
In each case, the effect of turning on a quantum symmetry is to reduce the
orbifold group to a nonanomalous subgroup (correlated to the choice
of $B$), precisely as advertised.

This example was also studied, from a different perspective,
in \cite[appendix A]{Robbins:2021lry}.

\subsection{Anomalous ${\mathbb Z}_2$ extended to
${\mathbb Z}_2 \times {\mathbb Z}_2$}

This example will be of a different form than previous examples
we have considered:  in this case, the extension itself will not
trivialize the anomaly.  Nevertheless, we will see that the anomaly
can be removed via a choice of discrete torsion.

Consider an anomalous orbifold $[X/{\mathbb Z}_2]$,
with anomaly $\alpha \in H^3({\mathbb Z}_2,U(1)) = {\mathbb Z}_2$, 
and extend the
orbifold group to $K \times G = {\mathbb Z}_2 \times {\mathbb Z}_2$,
with $K = {\mathbb Z}_2$ acting trivially on $X$:
\begin{equation}
1 \: \longrightarrow \: {\mathbb Z}_2 \: \longrightarrow \:
{\mathbb Z}_2 \times {\mathbb Z}_2 \: \stackrel{\pi}{\longrightarrow} \:
{\mathbb Z}_2 \: \longrightarrow \: 1.
\end{equation}
Since the extension splits, $d_2 B$ is trivial for any choice of 
quantum symmetry $B$, and so there is no choice of $B$ such that
$\alpha = d_2 B$ for any nontrivial $\alpha$.

Nevertheless, let us apply decomposition to various choice of
quantum symmetry $B$ and discrete torsion $\omega$.
The possible quantum symmetries are 
\begin{equation}
H^1(G, H^1(K,U(1))) \: = \: {\mathbb Z}_2,
\end{equation}
which we will enumerate via their action on the generator of
$G$, which we will denote $a$.  If we let $\omega$ denote the nontrivial
choice of discrete torsion in $H^2({\mathbb Z}_2 \times {\mathbb Z}_2,
U(1))$ then from e.g.~\cite[section 5.1]{Robbins:2021ylj},
we know that $\iota^* \omega = 0$ and $\beta(\omega) \neq 0$.
Applying the decomposition conjecture of
\cite{rsv1}, we find the quantum field
theory of $[X/\Gamma]_{B, \omega}$, as listed in table~\ref{table:ex:z2z2:z2}.

\begin{table}[h]
\begin{center}
\begin{tabular}{c|cc}
$B(a)$ & w/o d.t. & with d.t. \\ \hline
$+1$ & $\coprod_2 [X/{\mathbb Z}_2]$ & $X$ \\
$-1$ & $X$ & $\coprod_2 [X/{\mathbb Z}_2]$
\end{tabular}
\caption{Results for QFT($[X/\Gamma]_{B,\omega}$) for various choices of
$B$, $\omega$.  As $\beta(\omega) \neq 0$, the result of turning on
discrete torsion is essentially to exchange the results for the
two values of $B$. \label{table:ex:z2z2:z2}
}
\end{center}
\end{table}

First, table~\ref{table:ex:z2z2:z2} confirms that if there is no quantum
symmetry and no discrete torsion, then the anomaly is not resolved,
one gets instead two copies of the anomalous ${\mathbb Z}_2$ orbifold.
However, if either there is a quantum symmetry or there is discrete torsion,
the anomaly is resolved, despite the fact that in no case in this
example does $d_2 B = \alpha$.  (Intuitively, discrete torsion or nonzero $B$
in the
$\Gamma$ orbifold has the effect of canceling out some of the sectors appearing
in the $G$ orbifold.)  We therefore emphasize that the anomaly
resolution method we have primarily focused on elsewhere in this paper
is not the only way to resolve anomalies.

This example was also considered, from a different perspective,
in \cite[section 5.1.2]{Robbins:2021lry}.

\section{Conclusions}

In this paper we have applied decomposition to simplify
the anomaly-resolution program described in 
e.g.~\cite{Wang:2017loc,Bhardwaj:2017xup,Tachikawa:2017gyf}, by showing that
enlarging the orbifold group and turning on a quantum symmetry phase
is equivalent to (copies of) orbifolding by a nonanomalous subgroup of
the original orbifold group.

\section*{Acknowledgements}

We would like to thank R.~Donagi, T.~Pantev, R.~Szabo,
and Y.~Tachikawa for useful
conversations.  D.R. was partially supported by
NSF grant PHY-1820867.
E.S. was partially supported by NSF grant
PHY-2014086.

\appendix

\section{Notes on degree-three group cohomology}
\label{app:deg3-gpcohom}

In this section, to make this paper self-contained, we collect
pertinent results on $H^3({\mathbb Z}_2 \times {\mathbb Z}_2, U(1)) = 
({\mathbb Z}_2)^3$.

Explicit cocycles representing this group cohomology were given 
in \cite[equ'n (5.26)]{Robbins:2021lry}, which we repeat below.
Writing ${\mathbb Z}_2 \times {\mathbb Z}_2 = \{1, a, b, c\}$,
we can represent $\alpha \in H^3({\mathbb Z}_2 \times {\mathbb Z}_2,
U(1))$ as
\begin{equation}
\begin{matrix}
        \label{z2z2omega}
        \alpha(a,a,a)=\epsilon_a, & \alpha(a,a,b)=1, & \alpha(a,a,c)=\epsilon_a, 
& \alpha(a,b,a)=1, & \alpha(a,b,b)=1, 
\\ 
& \alpha(a,b,c)=1, & \alpha(a,c,a)=\epsilon_a, & \alpha(a,c,b)=1, 
& \alpha(a,c,c)=\epsilon_a, 
\\ 
\alpha(b,a,a)=1, & \alpha(b,a,b)=1, & \alpha(b,a,c)=1, 
& \alpha(b,b,a)=\epsilon_a\epsilon_b\epsilon_c, & \alpha(b,b,b)=\epsilon_b, 
\\ 
& \alpha(b,b,c)=\epsilon_a\epsilon_c, 
& \alpha(b,c,a)=\epsilon_a\epsilon_b\epsilon_c, 
& \alpha(b,c,b)=\epsilon_b, & \alpha(b,c,c)=\epsilon_a\epsilon_c, 
\\ 
\alpha(c,a,a)=\epsilon_a, & \alpha(c,a,b)=1, & \alpha(c,a,c)=\epsilon_a, 
& \alpha(c,b,a)=\epsilon_a\epsilon_b\epsilon_c, 
& \alpha(c,b,b)=\epsilon_b, 
\\ 
& \alpha(c,b,c)=\epsilon_a\epsilon_c, 
& \alpha(c,c,a)=\epsilon_b\epsilon_c, 
& \alpha(c,c,b)=\epsilon_b, & \alpha(c,c,c)=\epsilon_c,
\end{matrix}
\end{equation}
where $\epsilon_{a,b,c}$ represent the cohomology class, each $\epsilon^2 = 1$,
and where if any index is $1$, $\alpha = 1$.

Now, of course, a given cocycle need not have the precise form above to
represent a given cohomology class, so below we list some invariants that
can be computed from a cocycle to determine the cohomology class (we assume that $\alpha$ is normalized, so that $\alpha(1,g,h)=\alpha(g,1,h)=\alpha(g,h,1)=1$ for all $g,h\in\Z_2\times\Z_2$):
\begin{eqnarray}
\alpha(a,a,a) & = & \epsilon_a, \label{eq:invt1}
\\
\alpha(b,b,b) & = & \epsilon_b,
\\
\alpha(c,c,c) & = & \epsilon_c,
\\
\alpha(a,b,a) \alpha(a,ab,a) & = & \epsilon_a,
\\
\alpha(b,c,b) \alpha(b,bc,b) & = & \epsilon_b,
\\
\alpha(c,a,c) \alpha(c,ac,c) & = & \epsilon_c. \label{eq:invt6}
\end{eqnarray}
(These were extracted from \cite[equ'n (3.2)]{Robbins:2021lry} and
\cite[equ'ns (5.36)-(5.41)]{Robbins:2021lry}.)
It is straightforward to check that the products of cocycles on the left
side of each equation are invariant under coboundaries.

\end{document}